\documentclass[aps,prl,supexrscriptaddress,twocolumn,
  floatfix,nofootinbib]{revtex4}
\usepackage{color,amsmath,amssymb,graphicx,tabularx,
  url,multirow,hyperref,booktabs,xspace}

\def\cls{\text{CL}_\text{s}}

\def\nj{N_\text{jets}}

\newcommand\one{\leavevmode\hbox{\small1\normalsize\kern-.33em1}}

\newcommand{\ope}{\mathcal{O}}

\newcommand{\met}{\slashchar{E}_T}

\newcommand{\gev}{\text{GeV}}
\newcommand{\tev}{\text{TeV}}

\newcommand{\ifb}{\text{fb}^{-1}}

\def\slashchar#1{\setbox0=\hbox{$#1$}           
   \dimen0=\wd0                                 
   \setbox1=\hbox{/} \dimen1=\wd1               
   \ifdim\dimen0>\dimen1                        
      \rlap{\hbox to \dimen0{\hfil/\hfil}}      
      #1                                        
   \else                                        
      \rlap{\hbox to \dimen1{\hfil$#1$\hfil}}   
      /                                         
   \fi}

\newcommand{\eg}{\textsl{e.g.}\;}

\newcommand{\be}{\begin{eqnarray*}}
\newcommand{\ee}{\end{eqnarray*}}

\newcommand{\bee}{\begin{eqnarray}}
\newcommand{\eee}{\end{eqnarray}}
\newcommand{\beeq}{\begin{equation}}
\newcommand{\eeeq}{\end{equation}}

\begin{document}

\title{LHC multijet events as a probe for anomalous dimension-six
  gluon interactions}

\author{Frank Krauss}
\affiliation{Institute for Particle Physics Phenomenology,
  Durham University, Durham DH1 3LE, United Kingdom}

\author{Silvan Kuttimalai}
\affiliation{SLAC National Accelerator Laboratory, Menlo Park, California 94025, USA}
\affiliation{Institute for Particle Physics Phenomenology,
  Durham University, Durham DH1 3LE, United Kingdom}

\author{Tilman Plehn}
\affiliation{Institut f\"ur Theoretische Physik,
  Universit\"at Heidelberg, 69120 Heidelberg, Germany}

\begin{abstract}
  Higher-dimensional multigluon interactions affect essentially all
  effective Lagrangian analyses at the LHC. We show that, contrary to
  common lore, such operators are best constrained in multijet
  production. Our limit on the corresponding new physics scale in the
  multi-TeV range exceeds the typical reach of global dimension-six
  Higgs and top analyses. This implies that the pure Yang-Mills
  operator can safely be neglected in almost all specific
  higher-dimensional analyses at Run~II.
\end{abstract}

\maketitle

With the first analyses of Run~II data of the LHC appearing, effective
Lagrangians~\cite{effective,buwy} are rapidly developing into the main
physics framework describing searches for physics beyond the standard
model.  Global analyses of Run~I and early Run~II data already exist
for the Higgs and electroweak gauge sectors~\cite{sfitter} and for the
top sector~\cite{top}, illustrating the power of this approach.  The
fact that essentially all production processes in all physics sectors
involve incoming gluons poses a major, unsolved challenge to all such
effective Lagrangian analyses: the pure Yang-Mills operator with
its corresponding Wilson coefficient
\begin{align}
c_G \ope_G &= \frac{g_s \, c_G}{\Lambda^2} \, f_{abc} 
         G_{a \nu}^\rho G_{b \lambda}^\nu G_{c \rho}^\lambda \notag \\
&\text{with} \quad 
G_a^{\rho \nu} 
= \partial^\rho G_a^\nu - \partial^\nu G_a^\rho - i g_s f_{abc} G^{b
  \rho} G^{c \nu}
\label{eq:g3}
\end{align}
will correlate all such analyses~\cite{sally,scott} and force us into
an unwieldy, if not unrealistic, global analysis of all LHC channels.
  The operator $D^\mu G_{\mu \nu} \, D_\rho G^{\rho \nu}$ can lead to
  similar effects, but it can be removed from our operator basis
  through equations of motion, mapping it to four-quark
  operators~\cite{cho_simmons}.
\smallskip

It is very well known that the contribution of $\ope_G$ to dijet
production in gluon-gluon or gluon-quark scattering does not interfere
with the standard model process~\cite{simmons}. Heavy quark production,
$gg \to t\bar{t}$ is an exception, and it can be used to constrain
$c_G/\Lambda^2$ at the Tevatron~\cite{cho_simmons}. However, the
operator $\ope_G$ is only one of many operators contributing to top
pair production, giving marginalized Run~I constraints of the order
$\Lambda/\sqrt{c_G} \gtrsim 850$~GeV~\cite{top}. Alternative, but less
powerful search strategies include four-jet production at
LEP~\cite{duff} and three-jet production at hadron
colliders~\cite{yael}, while the suggestion to constrain $\ope_G$ in a
Higgs analysis~\cite{wiebusch} lacks realism given the current reach
of such a Higgs analysis~\cite{sfitter}.\smallskip

In this letter we propose to search for effects of $\ope_G$ in a new
channel, namely multijet production which we analyze for up to six
hard jets. Our analytic understanding of inclusive and exclusive
multijet production processes has matured~\cite{staircase}, and we
can robustly and precisely simulate such processes~\cite{qcd_nlo}. In
this note we will rely on two well-controlled observables,
namely the (exclusive) number of jets $\nj$ and $S_T$, defined as the
scalar sum of jet transverse momenta plus any missing transverse
energy exceeding 50~GeV~\cite{CMS:2015iwr},
\begin{align}
S_T = \left( \sum_{j=1}^{\nj} E_{T,j} \right) + \left( \met > 50~\gev \right) \; .
\label{eq:s_t}
\end{align}
The two observables allow the separation of two-jet production from
events with a larger number of jets while simultaneously giving
a measure of the energy scale tested in the partonic process.

Two-jet production from partonic processes such as
$q\bar{q} \to q'\bar{q}'$ serves as an excellent probe of four-quark
effective operators. Because this topology carries little sensitivity
to $\ope_G$~\cite{simmons} we will impose the corresponding ATLAS
limits on four-quark operators~\cite{fourquark} in our multijet
analysis in order to limit the effect of these operators.\smallskip

Our effective Lagrangian hypothesis is defined by following the standard
approach of global effective Lagrangian analyses~\cite{sfitter,top} to
test the dimension-six Lagrangian only as a well-defined hypothesis.  
The effect of the corresponding dimension-six operators in generic
multijet signatures scales like $E^2/\Lambda^2$, but the wide available
energy range at the LHC sheds some doubt on the assumption that the effects
of dimension-eight operators are systematically suppressed compared to
dimension-six operators. We therefore treat the effects of higher-dimensional
operators as theoretical uncertainties in the matching procedure of a
given full model to the dimension-six Lagrangian~\cite{validity}.\bigskip

\begin{figure*}[t]
\includegraphics[width=0.46\textwidth]{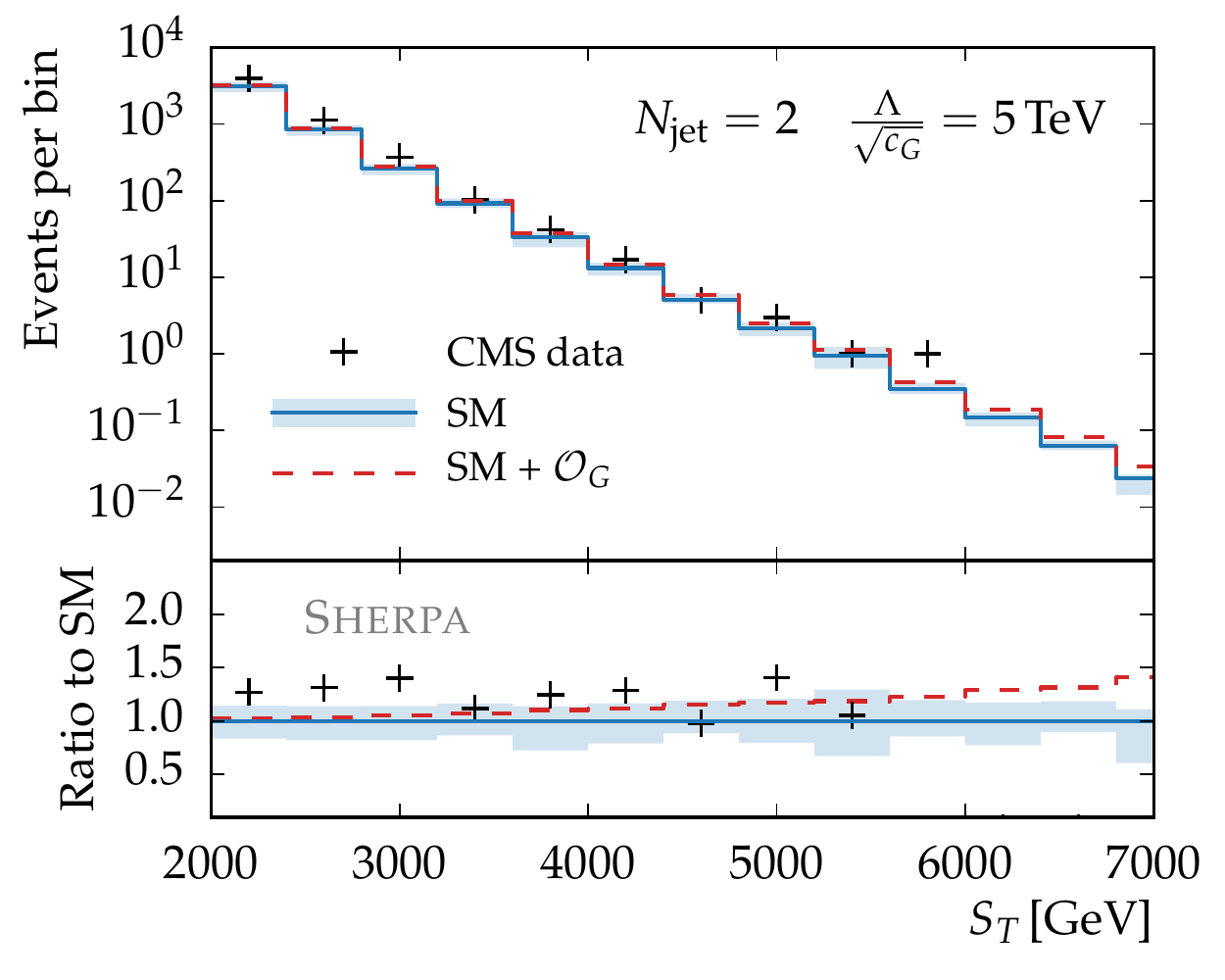}
\hfill
\includegraphics[width=0.46\textwidth]{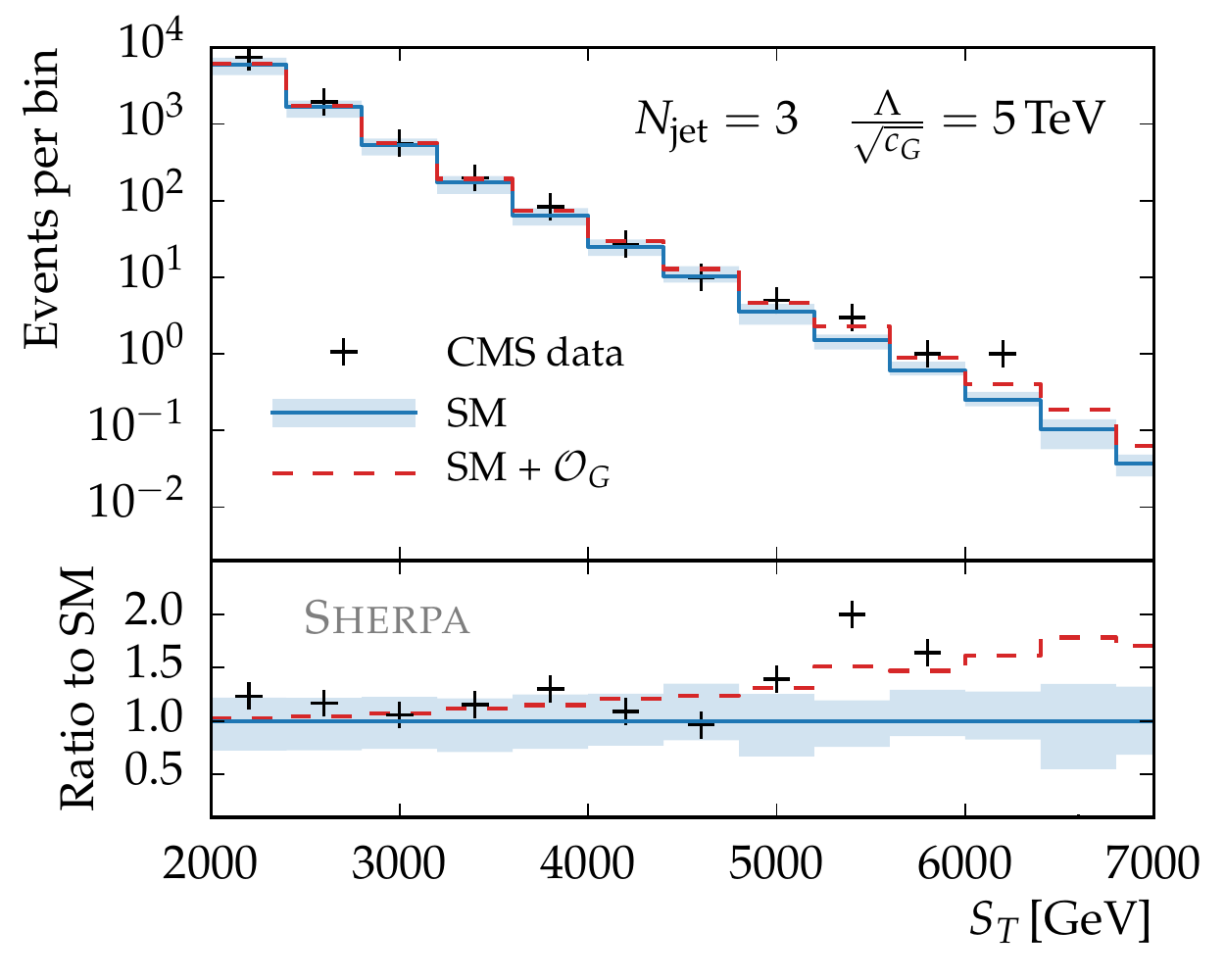}

\includegraphics[width=0.46\textwidth]{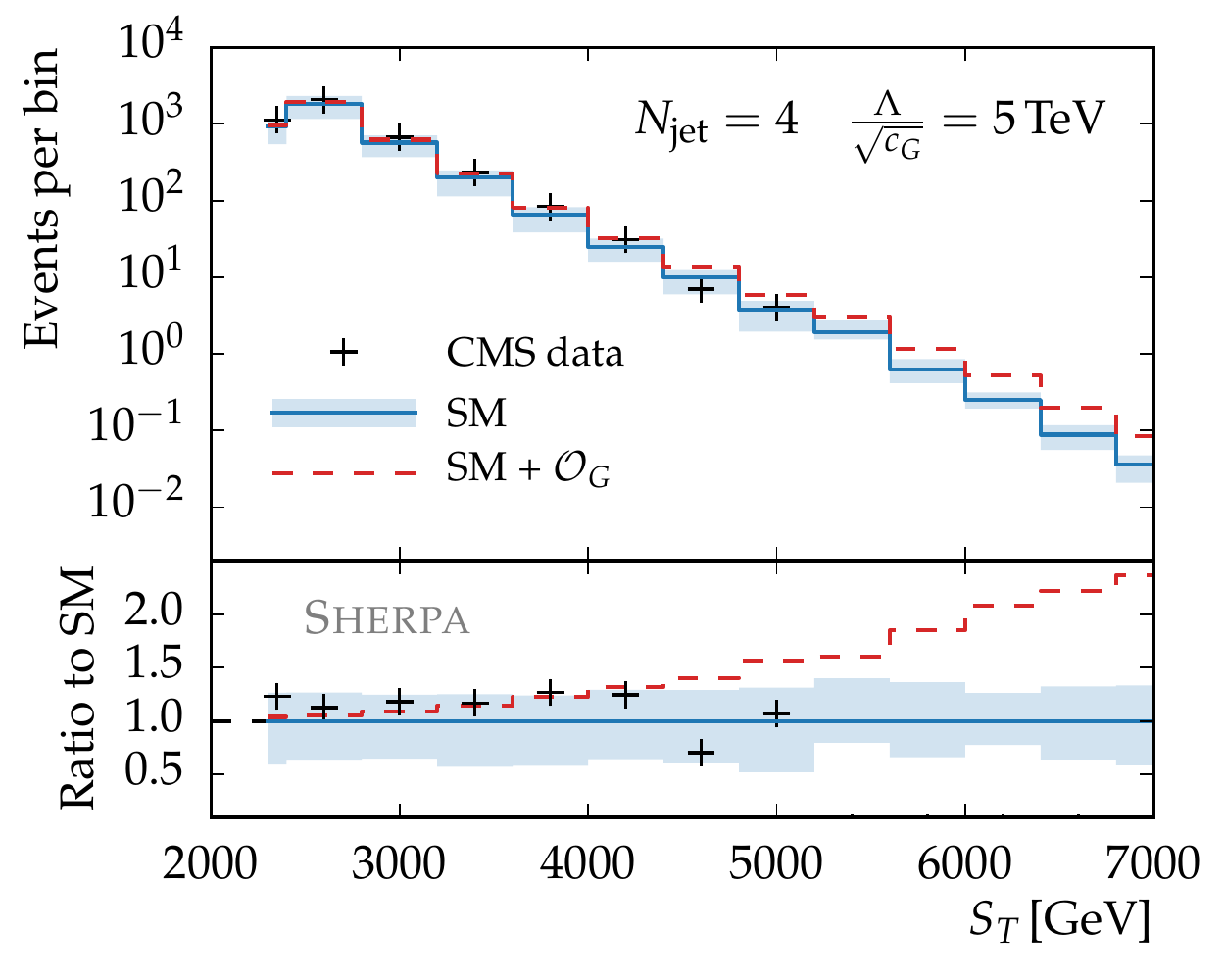}
\hfill
\includegraphics[width=0.46\textwidth]{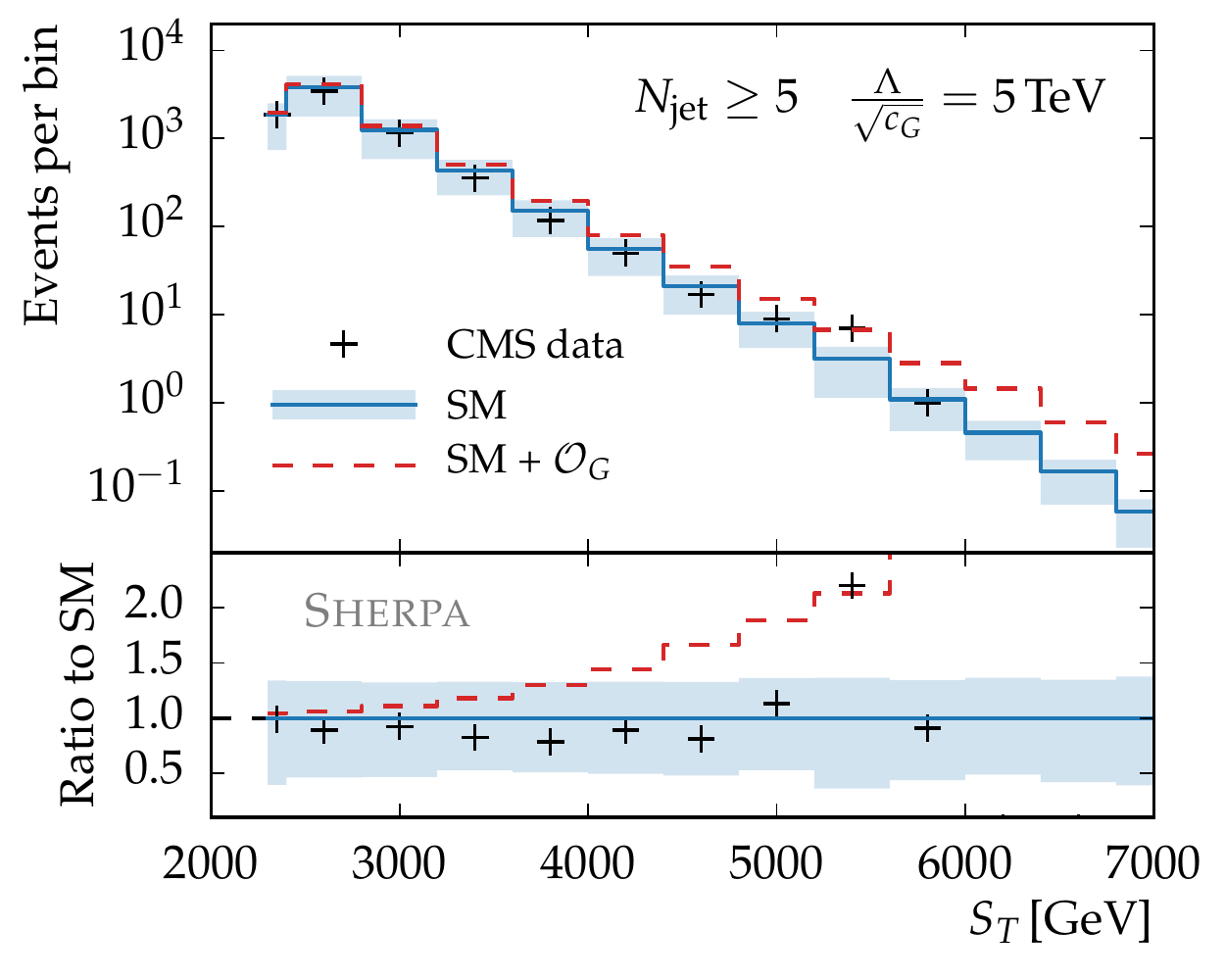}
\vspace*{-5mm}
\caption{$S_T$ distributions from CMS~\cite{CMS:2015iwr} in various
  bins of exclusive/inclusive jet multiplicity $\nj$, compared to our
  multijet-merged signal and background predictions including
  perturbative uncertainties.}
\label{fig:ht_sm}
\end{figure*}

\underline{Multijet signature} --- Our analysis of the dimension-six
QCD Lagrangian is based on a CMS search for extradimensional black
holes~\cite{CMS:2015iwr}, which to date is the only published 13~TeV
analysis based on a sizeable data set and extending to a large number
of jets without requiring any additional particles in the final state.
Obviously, dedicated ATLAS or CMS analyses of multijet production in
the light of dimension-six operators will improve upon our results.
The background is completely dominated by QCD jet production, so
just as in the original analysis we neglect non-QCD backgrounds.

For a robust description of the high-multiplicity QCD jet backgrounds,
we employ CKKW multijet merging within \textsc{Sherpa}~\cite{ckkw,sherpa},
with next-to-leading-order matrix elements for dijet production
and leading-order matrix elements for up to six jets in the final state.
Our nominal choice for the factorization and renormalization scales is
determined by a backwards clustering procedure and the scale choice
$\sqrt{2} \mu_{r,f}^2 =1/(s^{-1}+t^{-1}+u^{-1})$ for the $2\to 2$ core
process~\cite{ckkw}.\smallskip

As shown in Fig.~\ref{fig:ht_sm}, the observed $S_T$ distributions are
accurately described by our SM simulations. We estimate perturbative
uncertainties through independent variation of both scales by a
factor of two around the nominal values, omitting combinations
where one scale is varied upwards and the other one downwards to
avoid large logarithms.  All differences between data and
the SM simulation are within the estimated perturbative uncertainties.
The minimal tension in the exclusive two-jet bin at low $S_T$ only
occurs after translating the original inclusive results into jet-exclusive
distributions. They will not affect our analysis of the multijet
rates and our constraints on higher-dimensional operators contributing
to this process.\smallskip

Our signal simulations including the operator $\ope_G$ are based on an
implementation of the dimension-six operator of Eq.\eqref{eq:g3} in
\textsc{FeynRules}~\cite{feynrules}. We employ the \textsc{Ufo} output format in order
to facilitate event generation with \textsc{Sherpa} and its matrix element
generator \textsc{Comix}~\cite{ufo,sherpa_bsm}. For the purpose of implementing
the new exotic color structures that appear in the Feynman rules of
the dimension-six operator, a code generator module for arbitrary color
structures was implemented in \textsc{Sherpa}. This feature will become
publicly available along with the next \textsc{Sherpa} release. The automatic
generation of arbitrary Lorentz structures using \textsc{Sherpa} is described
in Ref.~\cite{sherpa_bsm}.

\begin{figure}[t]
\includegraphics[width=0.46\textwidth]{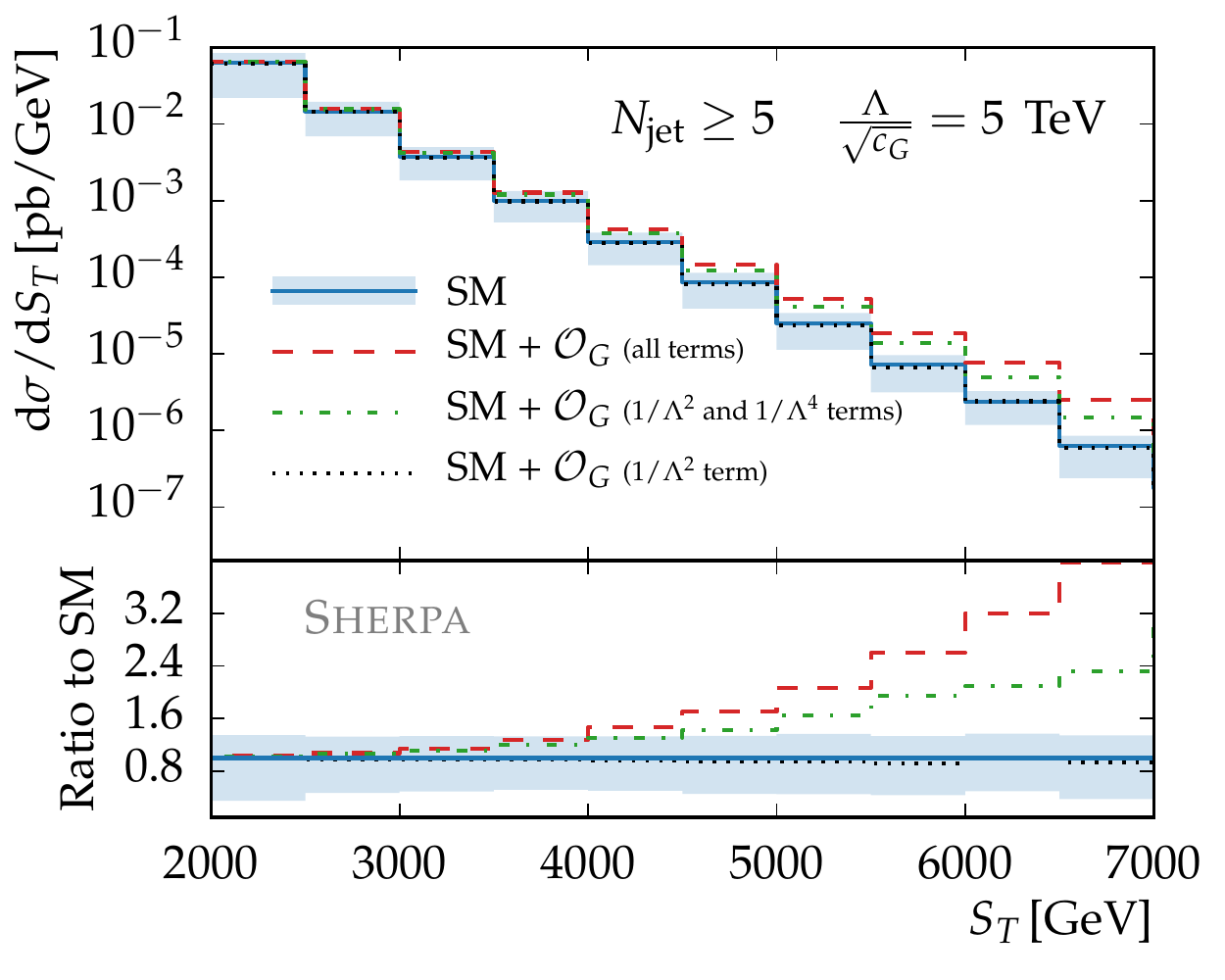}
\vspace*{-5mm}
\caption{Effect of multiple occurrences of the dimension-six Yang-Mills
  operator in the multijet matrix elements.}
\label{fig:multiple}
\end{figure}

\begin{figure}[t]
\includegraphics[width=0.46\textwidth]{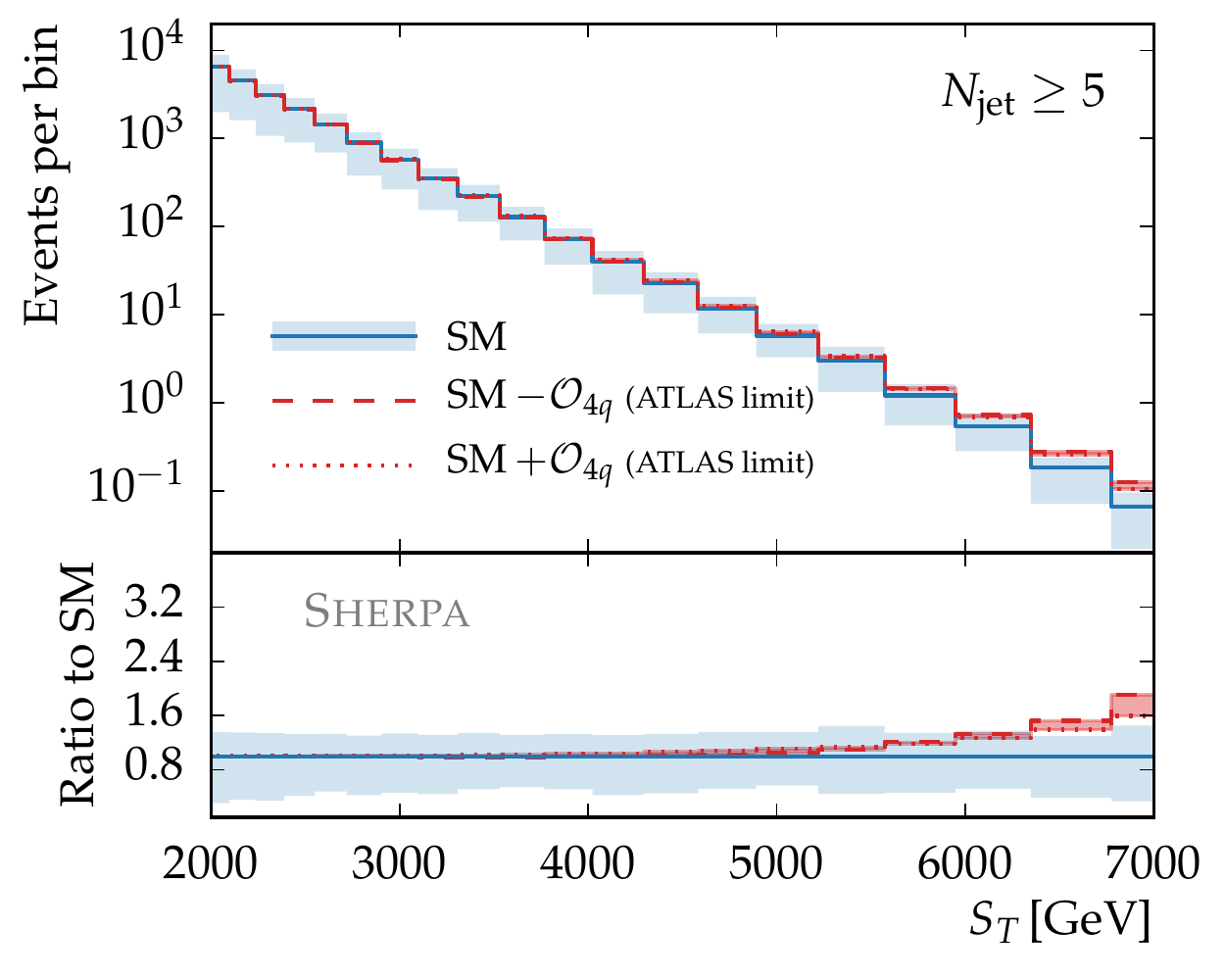}
\vspace*{-5mm}
\caption{Effect of effective four-quark operators in our signal
  region, with $\Lambda/\sqrt{c_{q4}}$ set to the lower limits
  obtained by ATLAS~\cite{fourquark}.}
\label{fig:fourquark}
\end{figure}

Just like the QCD background we compute the contributions of the
dimension-six operator of Eq.~\eqref{eq:g3} using CKKW multijet merging
techniques with leading-order matrix elements for up to five
jets~\cite{ckkw}. Formally, we can organize the effect of the higher
dimension contributions in terms of the scale suppression in the
multijet cross section. In this scheme, the leading interference
terms with SM diagrams are proportional to $1/\Lambda^2$, while the
dimension-six contributions squared contribute to $1/\Lambda^4$ or
higher, depending on the numerically relevant number of operator
insertions.

In Fig.~\ref{fig:multiple} we show the new physics effects in the
$S_T$ distribution for large jet multiplicities. The effects due to
interference terms proportional to $1/\Lambda^2$ are negligible
throughout the displayed range of $S_T$. Significant effects, however,
arise from terms of order $1/\Lambda^4$. This dominance of terms of
order $1/\Lambda^4$ over terms of order $1/\Lambda^2$ can also be
observed in top-pair production \cite{top}. For $S_T>\Lambda$, the
contributions due of terms of order $1/\Lambda^6$ and beyond
eventually become significant. This is to be expected, since
$S_T/\Lambda>1$ in this region, thus spoiling the parametric
suppression in $1/\Lambda$ and leading to a breakdown of the effective
field theory (EFT)
approach. This might lead to problems in matching our effective
Lagrangian results to a given full model. A standard solution to this
problem is to truncate the $S_T$ spectrum at $S_T=\Lambda$, thus
avoiding the kinematic region in which the EFT breaks down. Such a cut
is known to almost entirely remove the sensitivity to
higher-dimensional operators for example in Higgs
physics~\cite{sfitter}. The sensitivity of the analysis presented
here, however, is only very mildly affected by this cut, as will be
shown in what follows.\bigskip

\underline{Four-quark operator} --- While multijet production at the
LHC is dominated by gluon amplitudes, processes with quarks in the initial
and final states still lead to visible effects. These processes are
sensitive to the dimension-six contact interaction
\begin{align}
c_{q4} \ope_{q4} = \pm \frac{c_{q4}}{\Lambda^2} \sum_{q,q^\prime} 
\left( \bar q_L\gamma^\mu q_L \right) \; 
\left( \bar q^\prime_L\gamma^\mu q^\prime_L \right) \; .
\label{eq:q4}
\end{align}
While in principle the two operators in Eq.\eqref{eq:g3} and
Eq.\eqref{eq:q4} should be treated concurrently, we know from the
amplitude structure that the number of jets $\nj$ separates
their respective signal regions.  For the four-quark operator
the highest sensitivity can be obtained from two-jet correlations
and we therefore use the state-of-the-art result from the comprehensive,
multi-variate ATLAS analysis~\cite{fourquark}.  Being formulated
as an extension to resonance searches it does not include the
higher-dimensional gluon operator, and one should therefore use the
two-jet topology only.  There, the ATLAS analysis gives
\begin{align}
\frac{\Lambda}{\sqrt{c_{q4}}} > 4.79~...~6.8~\tev \; ,
\label{eq:atlas_limit}
\end{align}
in the conventions of Eq.\eqref{eq:q4} and depending on the assumed
sign of the Wilson coefficient.\smallskip

We estimate the impact of the four-quark operator on our Yang-Mills
analysis by computing its effect on multijet production. In
Fig.~\ref{fig:fourquark} we show the impact of the four-quark operator
within its allowed range of Eq.\eqref{eq:atlas_limit} on the multijet
signature. This result can be directly compared to the expected signal
from $\ope_G$, shown in Fig.~\ref{fig:multiple}.

Comparing the two effects on the high-energy tail of the $S_T$
distribution with an assumed new physics scale $\Lambda/\sqrt{c_G}
\lesssim 5$~TeV we confirm that the four-quark effects are strongly
suppressed. We find that the two effects only become comparable when
we increase the new physics scale in the Yang-Mills operator to
$\Lambda/\sqrt{c_G} \gtrsim 7$~TeV. \bigskip

\begin{figure*}[t]
\includegraphics[width=0.46\textwidth]{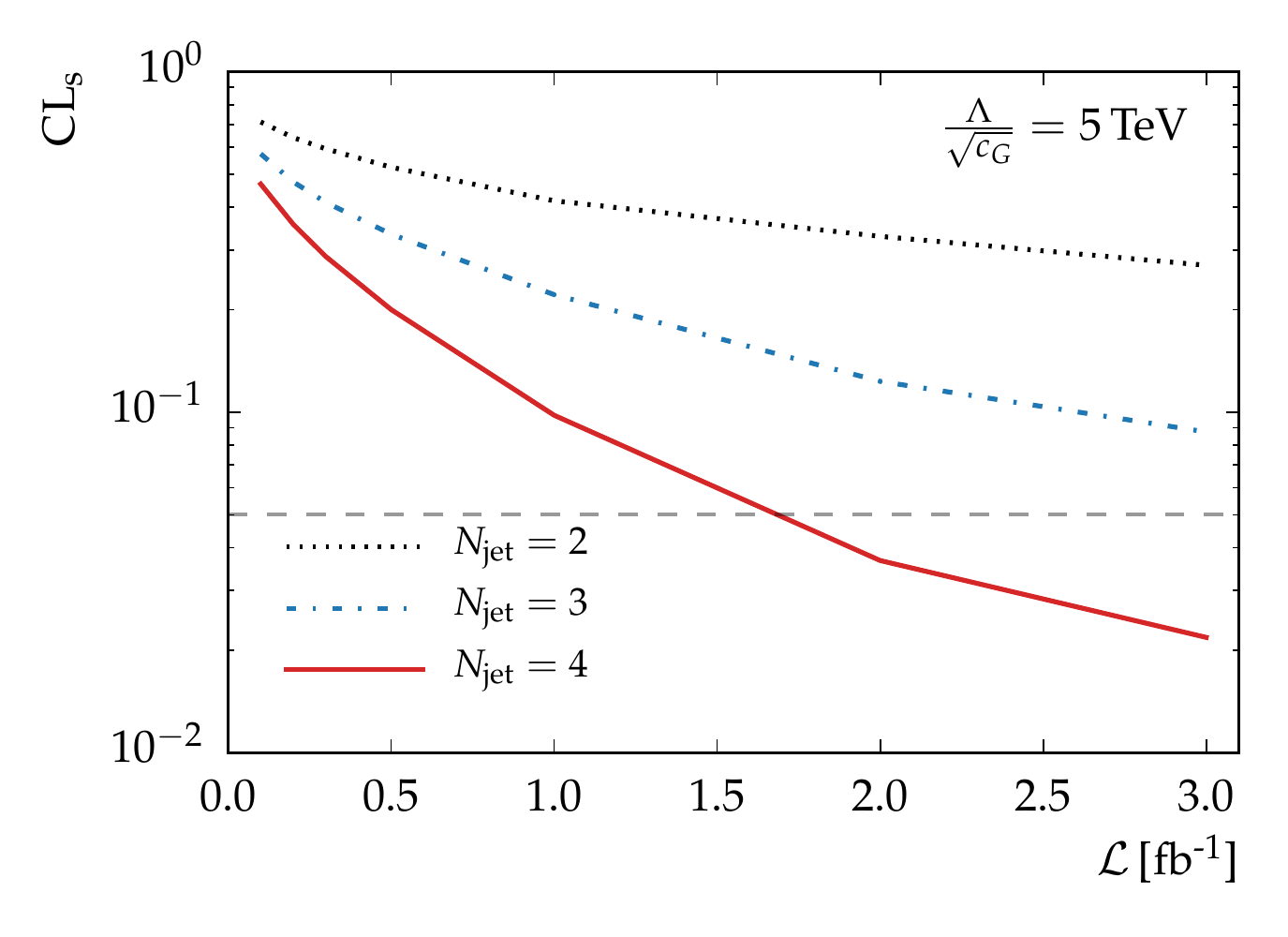}
\hfill
\includegraphics[width=0.46\textwidth]{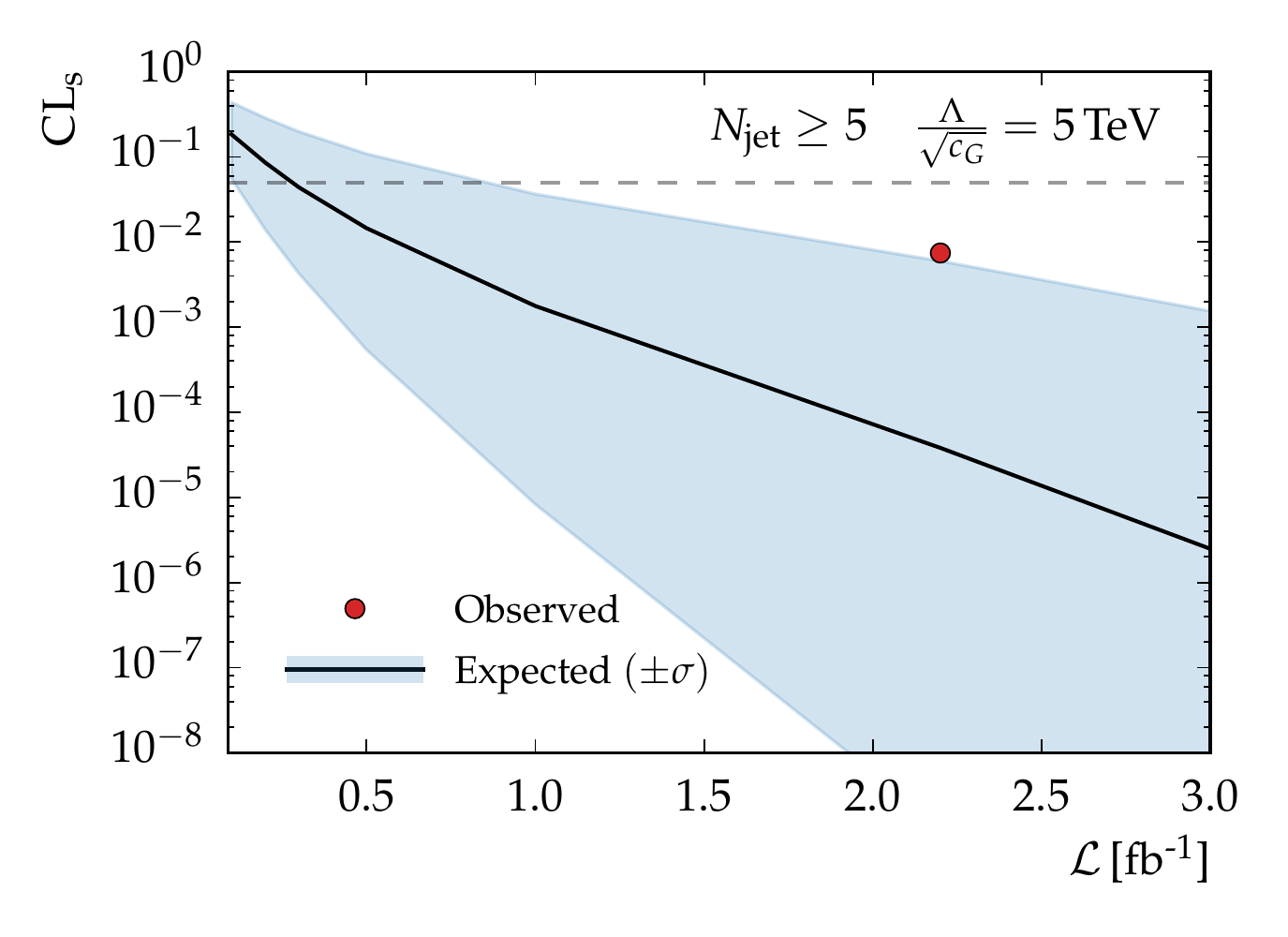}
\vspace*{-5mm}
\caption{Observed and expected signal confidence levels as a function
  of the integrated luminosity.  We show the expected results for
  fixed numbers of $\nj = 2.3.4$ (left) and for $\nj \ge 5$
  (right). An observed $\cls$ below the dashed line indicates a signal
  confidence below $5\%$ and allows for an exclusion of the
  dimension-six hypothesis.}
\label{fig:versus_lumi}
\end{figure*}

\underline{Multigluon operator limit} --- Finally, we can use the
$S_T$ distributions in bins of $\nj$ to constrain the Yang-Mills
operator $\ope_G$ in terms of a signal confidence $\cls$ as defined in
\cite{cls}. In the calculation of $\cls$ we take into account the
dominant systematic uncertainties, which are inherent in our background
predictions. In Fig.~\ref{fig:versus_lumi} we show the expected signal
confidence for $\Lambda/\sqrt{c_G} = 5$~TeV as a function of the
integrated luminosity collected at the LHC with $\sqrt{s}=13$~TeV. In
the left panel we see that indeed the sensitivity of the two-jet
topology is poor. This also confirms that adding the Yang-Mills
operator $\ope_G$ to the four-quark operator analysis of ATLAS will
not affect the limit shown in Eq.~\eqref{eq:atlas_limit}.

For higher jet multiplicities $\nj = 3,4$ the LHC reach slowly
increases, and we expect to rule out $\Lambda/\sqrt{c_G} < 5$~TeV
based on an integrated luminosity of less than $2~\ifb$. However, the
by far strongest constraints can be derived from the inclusive
five-jet sample, with a required luminosity well below $0.5~\ifb$ for
$\Lambda/\sqrt{c_G} = 5$~TeV.\smallskip

\begin{figure*}[b]
\includegraphics[width=0.46\textwidth]{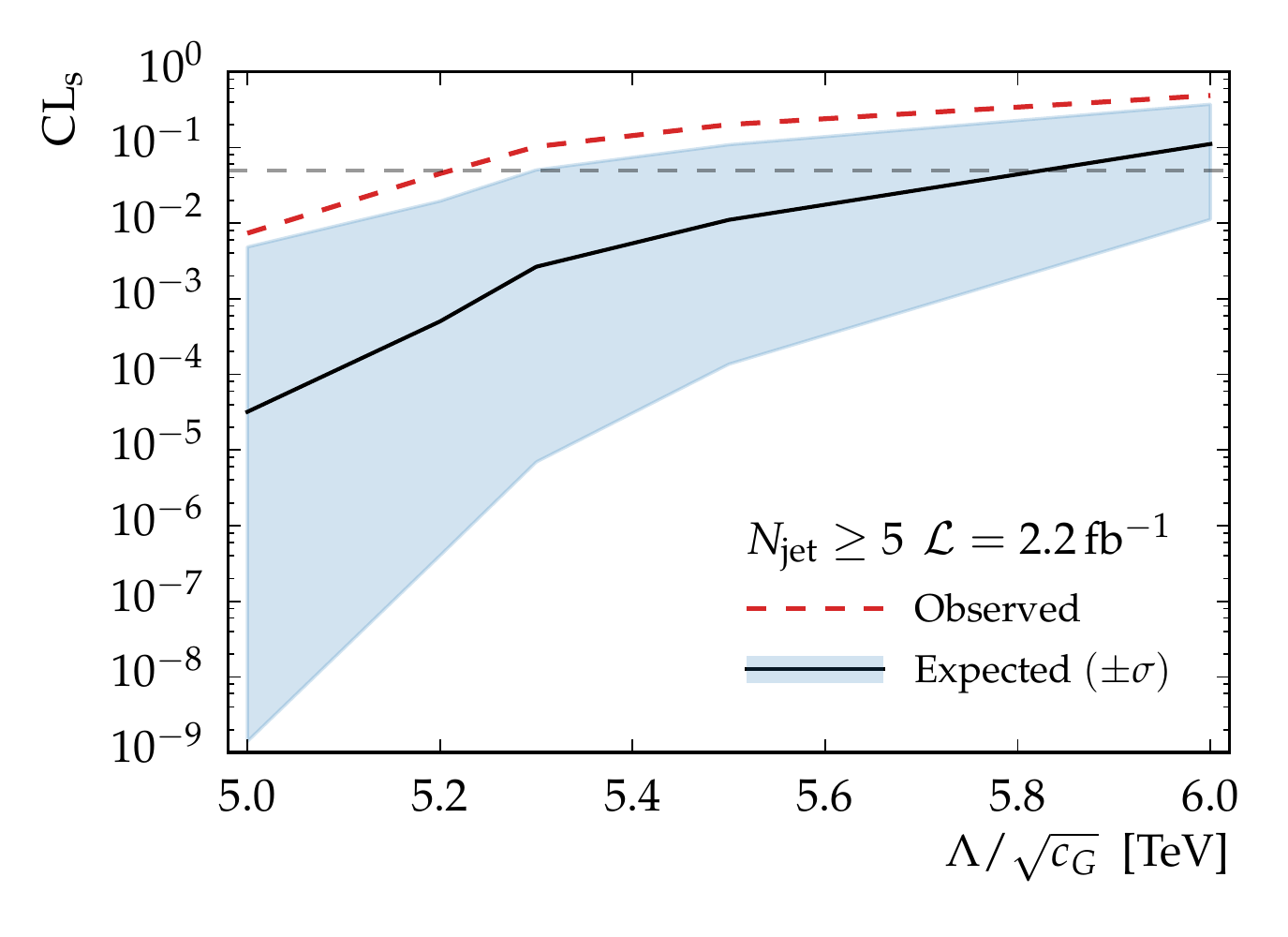}
\hfill
\includegraphics[width=0.46\textwidth]{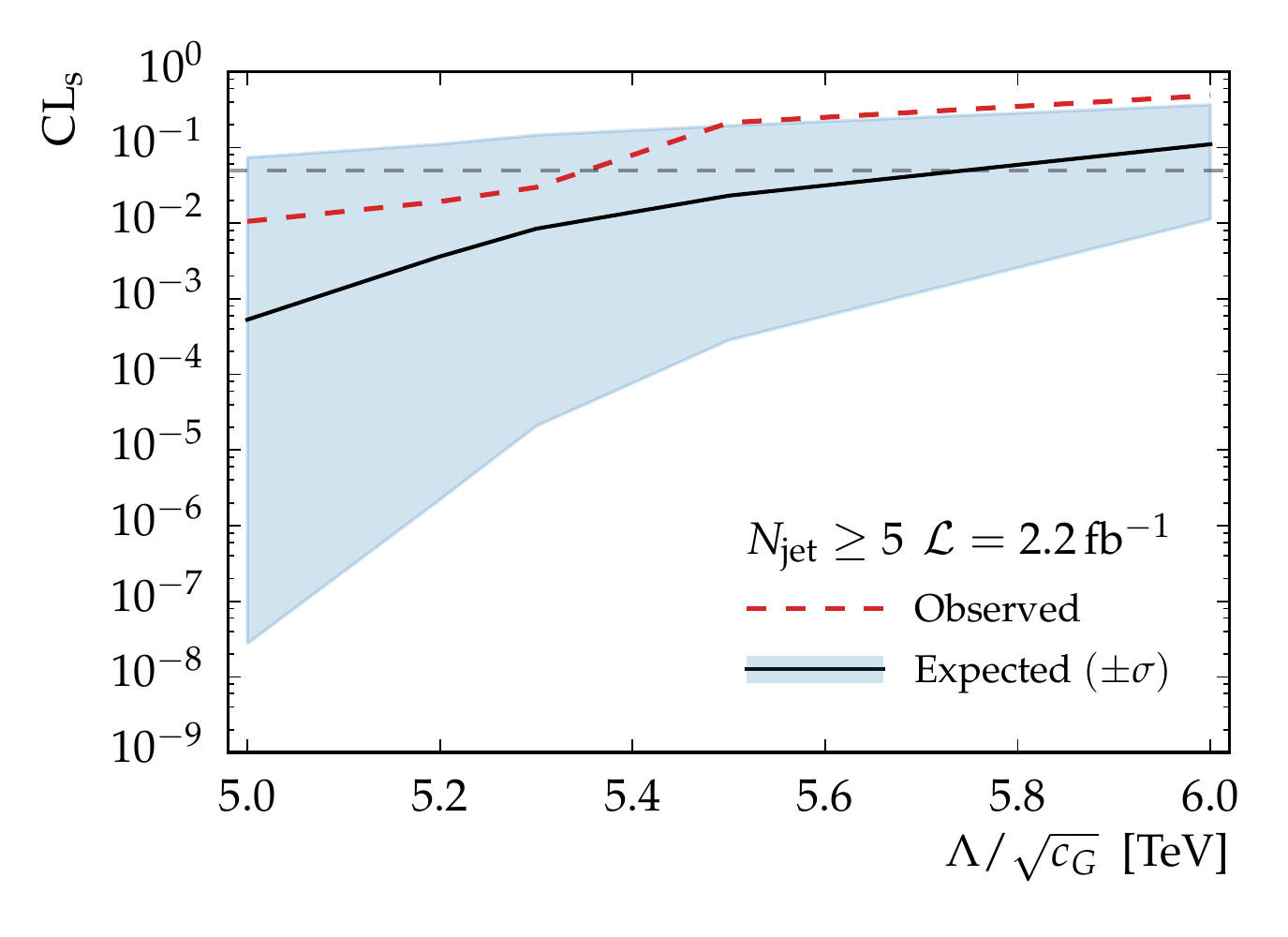}
\vspace*{-5mm}
\caption{Observed and expected signal confidence levels for the
  Yang-Mills operator $\ope_G$ as a function of $\Lambda/\sqrt{c_G}$.
  The results shown in the left take into account the full $S_T$
  distribution. The plot on the right-hand side shows the sensitivity
  of the analysis when truncating the distribution at $S_T=\Lambda$.}
\label{fig:versus_lambda}
\end{figure*}

In the conventions of Eq.\eqref{eq:g3} we find a limit on the
Yang-Mills operator $\ope_G$ of
\begin{alignat}{7}
\frac{\Lambda}{\sqrt{c_G}} &> 5.2~\tev 
\qquad &&\text{(observed)} \notag \\
\frac{\Lambda}{\sqrt{c_G}} &> 5.8~\tev 
\qquad &&\text{(expected),}
\end{alignat}
based on $\cls <5\,\%$ (see left panel of
Fig.~\ref{fig:versus_lambda}). The difference between expected and
observed limits corresponds to a deviation of just over one sigma and
is, in part, due to a slight excess in the data between $S_T=5~\tev$
and $S_T=6~\tev$, as shown in the lower right panel of
Fig.~\ref{fig:ht_sm}.

In Fig.~\ref{fig:versus_lambda}, we demonstrate that sensitivity of
our analysis is not an artifact of the very large new physics effects
in the region of $S_T>\Lambda$, where the applicability of the EFT is
questionable. We compare the expected and observed dependence of
$\cls$ on $\Lambda$ when taking into account all events and when taking into
account only events with $S_T<\Lambda$. As can be seen in this figure,
the expected sensitivity is only very mildly affected by this cut. The
observed limit on $\Lambda/\sqrt{c_G}$ is in fact stronger when
avoiding the region $S_T>\Lambda$, due to the presence of a slight
excess in the data above $S_T=5$\,TeV.\bigskip

\underline{Conclusions} --- The purely gluonic dimension-six operator
$\ope_G$ is known to be a major problem for all effective Lagrangian
analyses at Run~II. We show, for the first time, that it can very
effectively be constrained using multijet signatures at the LHC. Based on 
a CMS black-hole search with an integrated luminosity of
$2.2~\ifb$ at 13~TeV we find a limit $\Lambda/\sqrt{c_G} > 5.2$~TeV.
For an alternative definition $\ope_G = 1/\Lambda^2 \, f_{abc} G^3$
without the additional factor of $g_s$, we find
$\Lambda/\sqrt{c_G} > 4.7$~TeV. 

The effect of four-quark operators on our analysis can be fully
controlled by considering the two-jet and multijet signatures
separately. Our analysis demonstrates that possible effects of this
operator can be safely neglected in specific effective Lagrangian
analyses for example of the gauge, Higgs, or top sectors.\bigskip

\underline{Acknowledgments} --- We thank Stefan H\"oche for
interfacing our automatically generated exotic color structures in the
effective Lagrangian to \textsc{Comix} and acknowledge financial support by the
European Commission through the MCnetITN network
(PITN-GA-2012-315877).


\end{document}
